\documentclass[sigconf]{acmart}

\usepackage{color, colortbl}
\definecolor{Gray}{gray}{0.9}

\usepackage{booktabs} 

\usepackage{chngcntr}

\usepackage{enumitem}

\newcolumntype{P}[1]{>{\arraybackslash\footnotesize}p{#1}}
\newcolumntype{M}[1]{>{\centering\arraybackslash}m{#1}}
\newcommand*{\category}{\fontfamily{qcr}\selectfont}

\copyrightyear{2019}
\acmYear{2019} 
\setcopyright{iw3c2w3}
\acmConference[WWW '19]{Proceedings of the 2019 World Wide Web Conference}{May 13--17, 2019}{San Francisco, CA, USA}
\acmBooktitle{Proceedings of the 2019 World Wide Web Conference (WWW '19), May 13--17, 2019, San Francisco, CA, USA}
\acmPrice{}
\acmDOI{10.1145/3308558.3313716}
\acmISBN{978-1-4503-6674-8/19/05}

\begin{document}
\title{TurkScanner: Predicting the Hourly Wage of Microtasks}


\author{Susumu Saito}
\affiliation{%
  \institution{Waseda University}
  \city{Tokyo}
  \country{Japan}
  \postcode{169-8555}
}
\email{susumu@pcl.cs.waseda.ac.jp}

\author{Chun-Wei Chiang}
\affiliation{%
  \institution{West Virginia University}
  \city{Morgantown}
  \state{WV}
  \postcode{26505}
  \country{USA}}
\email{cc0051@mix.wvu.edu}

\author{Saiph Savage}
\affiliation{%
  \institution{Universidad Nacional Autonoma de Mexico (UNAM)}
  \city{Mexico City}
  \postcode{26505}
  \country{Mexico}}
\email{saiph.savage@mail.wvu.edu }

\author{Teppei Nakano}
\affiliation{%
  \institution{Waseda University}
  \city{Tokyo}
  \country{Japan}
  \postcode{169-0072}
}
\email{teppei@pcl.cs.waseda.ac.jp}

\author{Tetsunori Kobayashi}
\affiliation{%
  \institution{Waseda University}
  \city{Tokyo}
  \country{Japan}
  \postcode{169-0072}
}
\email{koba@pcl.cs.waseda.ac.jp}

\author{Jeffrey P. Bigham}
\affiliation{%
  \institution{Carnegie Mellon University}
  \city{Pittsburgh}
  \state{PA}
  \postcode{15213}
  \country{USA}
}
\email{jbigham@cs.cmu.edu}

\renewcommand{\shortauthors}{Susumu Saito, et al.}

\begin{abstract}
Workers in crowd markets struggle to earn a living.
One reason for this is that it is difficult for workers to accurately gauge the hourly wages of microtasks, and they consequently end up performing labor with little pay. In general, workers are provided with little information about tasks, and are left to rely on noisy signals, such as textual description of the task or rating of the requester. This study explores various computational methods for predicting the working times (and thus hourly wages) required for tasks based on data collected from other workers completing crowd work. We provide the following contributions. {\em (i)} A data collection method for gathering real-world training data on crowd-work tasks and the times required for workers to complete them; {\em (ii)} TurkScanner: a machine learning approach that predicts the necessary working time to complete a task (and can thus implicitly provide the expected hourly wage).
We collected 9,155 data records using a web browser extension installed by 84 Amazon Mechanical Turk workers, and explored the challenge of accurately recording working times both automatically and by asking workers.
TurkScanner was created using $\sim$150 derived features, and was able to predict the hourly wages of 69.6\% of all the tested microtasks within a 75\% error.
Directions for future research include observing the effects of tools on people's working practices, adapting this approach to a requester tool for better price setting, and predicting other elements of work ({\em e.g.}, the acceptance likelihood and worker task preferences.)
\end{abstract}

\ccsdesc[500]{Human-centered computing~Systems and tools for interaction design}

%
%


\keywords{crowdsourcing; Amazon Mechanical Turk; hourly wage}

\maketitle

\section{Introduction}

Workers on crowd platforms struggle to earn adequate wages \cite{irani2013turkopticon, mcinnis2016taking, ipeirotis2010analyzing}. This is problematic, given that one of the main motivators for crowd workers is to earn sufficient money to make a living 
\cite{berg2015income, brewer2016would, kuek2015global, martin2014being}. It is usually difficult for workers to earn higher than minimum wage, because crowd markets provide limited information on the offered microtasks. Workers struggle to gauge how much time a task will require, and to make informed judgements on whether tasks are worth their time. In this study, we introduce a machine learning approach to estimate the time required to complete a task and the approximate hourly wage of the task. 

Our work builds on recent work investigating crowd work wages. Callison-Burch et al. developed the CrowdWorkers browser extension \cite{callison2014crowd}, which records working times on Amazon Mechanical Turk (AMT). An analysis of this data revealed that the median hourly worker wage was only 2 USD/h \cite{hara2018data}. This is significantly lower than the U.S. minimum wage (7.25 USD/h). TurkBench leveraged historical records of completions of specific microtasks (called HITs on AMT) to suggest lucrative work to workers \cite{hanrahan2015turkbench}, but this was limited by which tasks had previously be seen. TurkBench could not estimate the hourly wage the first time that a particular type of task was performed.
New tasks are frequently posted in crowd marketplaces, and so it is important to be able to predict hourly wages for previously unseen microtasks. 

In this paper, we present TurkScanner, a machine learning approach for predicting the working times of microtasks to calculate their hourly wages based on previous logs of other workers for other tasks.
This allows workers to judge whether microtasks are worth doing, even for new tasks that no other workers have completed.

The first challenge we addressed in building TurkScanner was to collect reliable ground truth data on the working times of real tasks.
Estimating working times automatically is difficult, because we know that worker behavior patterns during microtasks and the motivations behind them are diverse \cite{kaplan2018striving}, such as visiting external websites to complete the tasks, taking breaks, or accepting a number of tasks and completing them in a row \cite{hara2018data}.
Rather than attempting to calculate a single working time, we collected three different times (two types of automatic recording and manual recording by the worker), along with the workers' a posteriori judgments on which were likely to be most accurate.
Our extension collected 9,155 data records of microtask submissions from 84 unique workers.
For each worker, we collected information on each of the tasks they completed, including the task (HIT) metadata and HTML content, the reputation of the requester, and the worker profile.
We aimed to collect all of the data that workers would view before actually completing a task.
Intuitively, we expected results such as ``HITs with longer times provided in their metadata may take longer,'' ``HITs posted by requesters with better ratings pay better,'' and ``HITs that included many input elements take longer to complete.''

Our second challenge was to predict working times (and thus hourly wages) for microtasks using a machine learning-based approach.
To the best of our knowledge, TurkScanner represents the first work to utilize machine learning to estimate the working times of microtasks.
We extracted $\sim$150 features from our data collected from AMT.
Our cross-validated results showed that TurkScanner achieved hourly wage predictions within a 75\% working-time error for 69.6\% of all the microtasks (and within a 100\% error for 84.3\%).

We conclude this paper by suggesting future research directions, to apply TurkScanner to support tools that empower both workers and requesters. We believe these types of machine learning mechanisms will enable a future in which crowd work becomes a more transparent and beneficial activity for all stakeholders.

\section{Related Work}

A wealth of previous literature has shown that workers are underpaid and unfairly treated in crowd markets \cite{horton2011condition, KatzAmazon,international2016non}.
Most crowdsourcing platforms allow requesters to freely create their microtasks and set their prices.
Requesters can also assess worker performances to control the quality of their answers, such as by screening workers' eligibility to work with qualifications \cite{hara2018data}, or by rejecting submitted tasks that do not meet their criteria \cite{bederson2011web}.
On the other hand, workers are usually provided with very limited information to select better microtasks \cite{irani2013turkopticon}.
For instance, most crowd markets only provide the task price, the name of the requester, a title, and a simple description. Such a lack of information makes it very challenging for workers to find good microtasks and requesters.
This power imbalance between requesters and workers limits workers' ability to ensure that they receive fair wages, and this results in many workers being paid below minimum wage \cite{irani2013turkopticon, mcinnis2016taking, ipeirotis2010analyzing,hitlin2016research,horton2010labor,irani2016stories,martin2014being}.
This problem was formalized after Hara et al. recently revealed that AMT workers earn a median hourly wage of only 2 USD/h, based on their data-driven analysis \cite{hara2018data}.
This surprising fact clearly emphasizes the necessity of tools to help crowd workers earn better wages.

Several researchers and practitioners have explored approaches to provide workers with better opportunities to obtain information on microtasks when selecting tasks \cite{mcinnis2016taking}. 
There are several web-based platforms on which workers communicate with each other about microtasks and requesters.
For instance, Turkopticon \cite{irani2013turkopticon} is an online forum and a tool where workers post reputation scores of requesters concerning several criteria, as well as free comments.
Among other support tools, Turkopticon is highly appreciated and actively utilized by workers \cite{kaplan2018striving}.
Many workers also join online social communities, such as Turker Nation \cite{turkernation} and MTurk Crowd \cite{mturkcrowd}. On these community websites, workers not only share reputations, but also recommend specific tasks and introduce tools and skills utilized by expert workers.
Web tools are also available to workers to directly facilitate crowd work within the crowd market \cite{chiang2018crowd}.
CrowdWorkers \cite{callison2014crowd} is a web browser extension for AMT workers, which collects microtask submission logs and visualizes the average hourly wage for each posted task.
TurkBench \cite{hanrahan2015turkbench} is another web platform for AMT workers, which renders a personalized, adaptive working schedule based on collected task logs, to suggest the most lucrative microtasks.

Like other tools, we aim to help workers select better microtasks.
More specifically, the goal of TurkScanner is to ``predict'' the hourly wages of new unseen microtasks, based on other tasks completed by other workers, rather than by calculations based on worker records for the same task.
We not only contribute the prediction algorithm, but also address the challenge of defining the ``working time'' of a microtask, so that we can calculate hourly wage based on this, which has not been sufficiently discussed in prior work.

\section{Methods}

In this section, we first describe our approach to defining the ``working time'' of a microtask.
Next, we explain our browser extension for data collection, designed based on our working time definitions.
We then explain the features of the dataset and its statistical analysis results.
Finally, we present the details of TurkScanner, which predicts the working times of microtasks and their hourly wages.

\subsection{Defining Working Time}

The ``working time'' refers to the amount of time required for workers to complete particular tasks.
To predict the working time using machine learning, it is essential to collect real-world data on actual working times.
However, measuring the working times is highly challenging.
We know that various worker behavior patterns exist, which prevents us from stably obtaining the correct working time using simple methods. During microtasks, workers often browse external websites in other tabs (either related or unrelated to the task), take breaks, work on multiple tasks in parallel, and so on.

In our approach, we recorded three different types of working time for every microtask.
Each of these has unique advantages and disadvantages.
When workers submitted microtasks, they were subsequently asked to choose one of the working times to finally label the collected data with.
The options were as follows:
\begin{itemize}[leftmargin=13pt,noitemsep,topsep=4pt]
    \item \textbf{TIME\_ALL}. Automatic recording method, consisting of the time from when workers accept the task until they finish it. \textit{Pros}: This is the most reliable working time when a worker immediately starts and completes a task without taking a break. \textit{Cons}: All of the time during which the worker is distracted (by checking emails, getting coffee, etc.) is counted.
    \item \textbf{TIME\_FOCUS}. An automatic recording method similar to TIME\_\\ALL, but only recording the time during which the microtask page tab is in focus. \textit{Pros}: This is a reliable working time even if a worker is distracted by task-irrelevant content in other tabs. \textit{Cons}: This approach cannot count the working time for microtasks that guide workers to other tabs at which the actual task resides (\textit{e.g.,} surveys or Google searches).
    \item \textbf{TIME\_BTN}. A manual recording method carried out by workers themselves, by toggling buttons to indicate when they are in the working status. \textit{Pros}: This is the most accurate working time when a worker utilizes the button correctly, covering all possible worker behavior patterns. \textit{Cons}: This approach is vulnerable to human error (\textit{e.g.,} spams).
\end{itemize}
For workers' final decisions, we also introduced the following working-time type as a fourth option:
\begin{itemize}[leftmargin=13pt,noitemsep,topsep=4pt]
    \item \textbf{TIME\_CUSTOM}. Manual input for the working time by workers in the case that none of the above three options seem to be correct. \textit{Pros}: This can provide workers with a last-resort option, to label the correct working time when all recording methods failed. \textit{Cons}: Errors may be present in workers' answers.
\end{itemize}

We proposed a few hypotheses concerning workers' judgments of the working time.
We expected the most dominant choice would be TIME\_BTN, because we anticipated that most workers would browse external websites or temporarily leave their computers for diverse reasons, and that the working time manually recorded by themselves would look more reliable in most cases.
Next, we supposed that TIME\_ALL would be the second most dominant choice.
However, we expected that this would still be chosen less than TIME\_BTN, because there are a certain number of workers who interleave tasks on crowdsourcing platforms \cite{kaplan2018striving}, such that the working time cannot be correctly measured by this approach.
Finally, we expected that TIME\_FOCUS and TIME\_CUSTOM would be chosen considerably less, but would still be useful when workers accidentally ignored the recording button.

\subsection{Data Collection With Web Browser Script}

We developed a Google Chrome extension for data collection, which crawls data from every completed microtask together with the actual working time. 
The extension records the working time automatically, provides AMT workers with a button for manually recording the working time, and asks workers to select the best choice for the working time with which to finally label the data.

\subsubsection{Worker Recruitment with Pre-Survey}

To recruit AMT workers, we prepared a pre-survey that asks for worker profile information ({\em e.g.}, gender, age, country, household income, worker experience, and worker hours per week). 
After the survey, we asked workers to install our extension according to instructions and a URL for the installation page.
After installation, each participant was given a unique ``installation code,'' to be copy-and-pasted back into the HIT to verify that they both installed the extension and completed the task.
The survey took about 4 min to complete, including the extension installation.
We paid workers 0.60 USD (\textit{i.e.,} an expected 9 USD/h) to complete the pre-survey and correctly paste the installation code.

\subsubsection{HIT Crawling using Chrome Extension}

When workers finished installing the extension, they were ready to begin data collection. 
Workers were asked to work on microtasks as usual.
The extensions collected data for up to 10 days after being installed, yet allowing the workers to uninstall it at any time (the workers were paid in proportion to their contribution until uninstallation). 
The data collection proceeded according to the three following steps (a), (b), and (c) (with workers paid a 5-cent bonus for completing each of (b) and (c) for each a HIT):

{\fontfamily{qhv}\selectfont\footnotesize
\begin{table}%
\def\arraystretch{1.3}
\caption{List of input features parsed from the collected data. The features consist of three categories and 12 sub-categories. The parenthesized numbers in bold text represent the feature dimension sizes.}
\label{tab:features}
\begin{center}
\begin{tabular}{P{0.5cm}P{0.3cm}p{6.6cm}}
  \toprule
  \rowcolor{Gray}
  \multicolumn{3}{l}{\textbf{HIT (71)} -- HIT-relevant information}\\
  META
  & \textbf{(3)}
  & HIT metadata set by requesters (\textit{e.g.,} reward, \# of available HITs)
  \\
  TMPL
  & \textbf{(11)}
  & HIT templates natively provided by AMT as of October 2018 (one-hot vector)
  \\
  URL
  & \textbf{(6)}
  & URL counts per content type included in a HIT page (\textit{e.g.,} anchor links, images, audios)
  \\
  INP
  & \textbf{(18)}
  & Input tag counts per type included in a HIT page (``type'' attributes such as \textit{radio} and \textit{text})
  \\
  TXT
  & \textbf{(1)}
  & Visible word counts in a HIT page
  \\
  KW
  & \textbf{(32)}
  & Task-relevant keyword occurrence in either HIT title, HIT description, or a HIT page (keywords such as ``survey'' and ''summarize'' arbitrary extracted by the authors)
  \\
  \rowcolor{Gray}
  \multicolumn{3}{l}{\textbf{WKR (28)} -- Worker-relevant information}\\
  PRFL
  & \textbf{(16)}
  & Worker profile information collected in pre-surveys (\textit{e.g.,} age, worker experience in years, est. hourly wage)
  \\
  EXT
  & \textbf{(8)}
  & AMT worker helper extension tools installed in web browser (\textit{e.g.,} CrowdWorkers \cite{callison2014crowd}, MTurk Suite \cite{mturksuite})
  \\
  HIST
  & \textbf{(4)}
  & Working history information as workers (\textit{e.g.,} approval rate, total earnings, \# of HIT submission in a HIT group) 
  \\
  \rowcolor{Gray}
  \multicolumn{3}{l}{\textbf{REQ (49)} -- Requester reputation information} \\
  TO
  & \textbf{(7)}
  & Turkopticon \cite{irani2013turkopticon} (\textit{e.g.,} average of 5-point scale requester evaluation of generosity, promptness, etc.)
  \\
  TO2
  & \textbf{(34)}
  & Turkopticon 2 \cite{irani2016stories} (\textit{e.g.,} average of 5-point scale HIT evaluation of recommendability, communicativeness, etc.)
  \\
  TV
  & \textbf{(8)}
  & TurkerView \cite{turkerview} (\textit{e.g.,} requester reputation for hourly wage settings, \# of reviews, etc.)
  \\
  \bottomrule
\end{tabular}
\end{center}
\bigskip\centering
\end{table}
}

\textbf{(a) Background data scraping.}
Once workers installed the extension, they were asked to select and work on any HIT as they would usually do on the platform.
The extension recorded the following data features, as shown in Table \ref{tab:features}.
When a worker visited a HIT page, the extension scraped HIT-relevant information ({\category HIT}), such as the HIT metadata and HTML contents of the HIT page (\textit{i.e.,} the text/media content information) that would possibly define task goals and required interactions in the task. 
In addition, the extension crawled worker-related information ({\category WKR}) from the worker dashboard and the list of installed AMT-relevant extensions once per day.
{\category WKR} would also be important, because its features would represent worker capabilities for microtasks \cite{rzeszotarski2011instrumenting}, such as their skill levels and the learning-curve effects \cite{yelle1979learning}.
Our extension also obtained requester-relevant information ({\category REQ)} concerning the reputations of requesters, provided via RESTful APIs in some third-party platforms ({\em i.e.}, Turkopticon \cite{irani_silberman_2013}, Turkopticon 2 \cite{irani_silberman_2017}, and TurkerView \cite{turkerview}).
This would help to understand how reasonably the microtask would be paid based on the requester ratings.


\begin{figure}
\includegraphics[width=8.5cm]{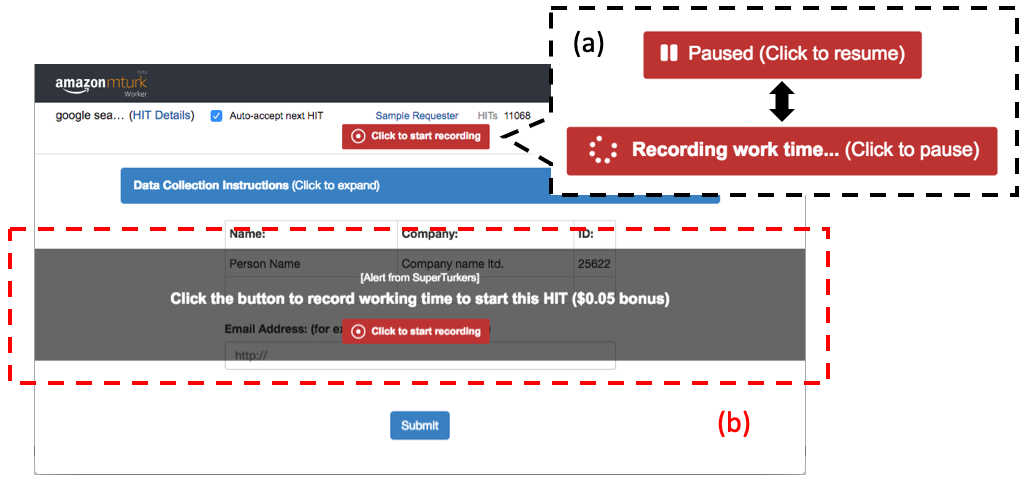}
\caption{Interface to record TIME\_BTN.
(a) The button at the top of the HIT page can be toggled to pause/resume recording working time.
(b) A black screen is rendered over the HIT at the beginning as a reminder workers to start the timer.}
\label{fig:record_button}
\end{figure}

\textbf{(b) Manual working time recording.}
While working on HITs, workers were asked to manually record the working time they spent on each HIT.
When workers accepted HITs, a red button was shown on the project details bar located at the top of a HIT page (Figure \ref{fig:record_button}a).
Workers recorded times at which they paused and resumed working on microtasks by toggling the button.
The two following additional features were implemented to remind workers to click the button when they start working on the HIT.
First, workers were prompted with a black screen rendered over the page (Figure \ref{fig:record_button}b).
This screen disappeared immediately once they clicked the button.
This made it less likely for workers to complete a HIT without dismissing the alert.
Second, a red frame was displayed around the window while the button was in the recording state, to assist workers in recognizing the recording state.
When workers interleaved HITs, ({\em i.e.}, worked on multiple HIT pages in multiple tabs simultaneously, by moving back and forth), none of the HITs would record their working times simultaneously, because we assumed that perfect multi-tasking of HITs is not possible.

\textbf{(c) Post-HIT survey.}
As soon as workers submitted a HIT, they were prompted with a "Post-HIT survey" in a popup window and they were asked to select the most accurate working time from the various methods with which to label HIT records.
The survey suggested four options for the HIT working time: TIME\_ALL, TIME\_FOCUS, TIME\_BTN, and TIME\_CUSTOM (see Section 3.1 for details).
Workers were required to input working time, in the form of X minutes and X seconds, only when they chose TIME\_CUSTOM.
The pop up window disappeared immediately after workers submitted their choices by clicking a ``Submit'' button. 

%
%
\subsection{Data Description}

Data collection was conducted for 10 days during late October 2018.
Our task dataset consisted of 9,155 HIT submission records collected by 84 unique workers.
The recorded HITs belonged to 1,641 unique HIT groups, posted by 998 unique requesters.
On average, workers contributed for 6.5 days (SD = 3.5; Median = 8.1) and worked on 109 HITs (Min = 1; Max = 1,958; SD = 238.1; Median = 34).

\textbf{Data Cleaning.}
To guarantee the quality of the dataset, we filtered out the following HIT submission records:
\textit{(i)} All HIT records of spam workers who used automated scripts (N=230);
\textit{(ii)} HIT records for which the reported working time was abnormally short or long (N=213);
\textit{(iii)} A large part of all the HIT records in the top three most-submitted HIT groups, keeping only as many HIT records as in the fourth largest HIT group, in order to reduce bias in the dataset (N=1,104).
After the data cleaning, 7,608 (83.1\%) HITs remained from 83 unique workers, for 1,587 HIT groups posted by 977 unique requesters.
This dataset was utilized for the analysis in the remainder of the paper.

\begin{figure}
\includegraphics[width=8.5cm]{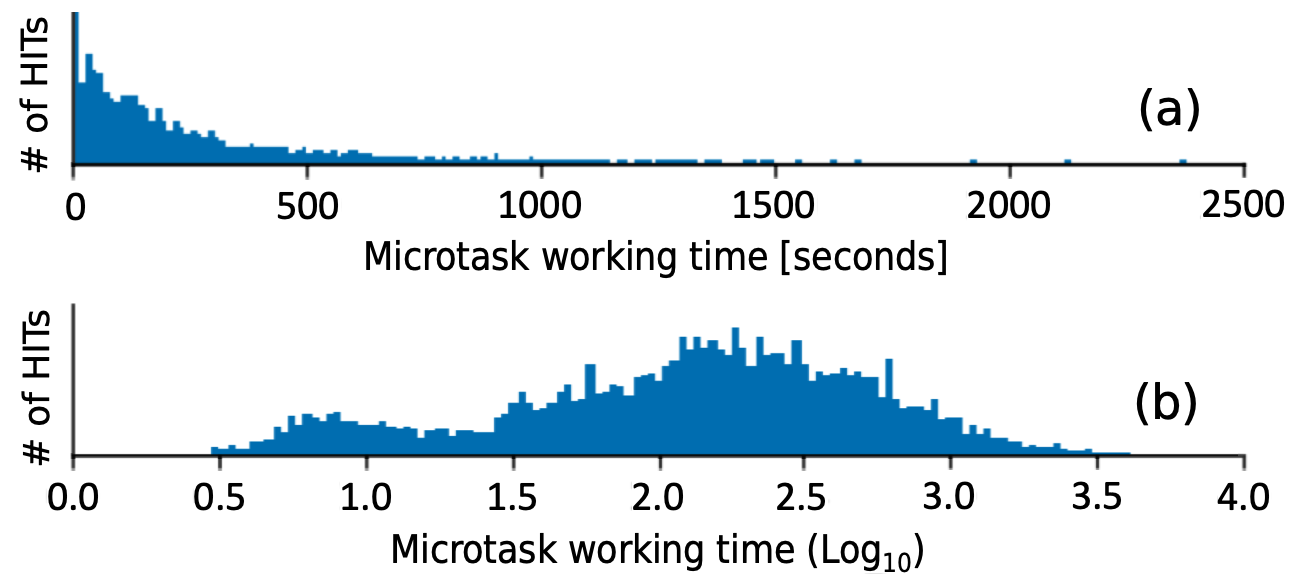}
\caption{
Working time distribution of microtasks in the dataset.
(a) Working time in seconds, which admits a long-tail distribution.
(b) $log_{10}$ working time, which can suppress the prediction error due to outliers.
}
\label{fig:working_time_dist}
\end{figure}

\textbf{Working Time Labels.}
On average, the reported working time was 277.9 s (SD = 380.2; Median = 148.3).
As in Figure \ref{fig:working_time_dist}(a), the distribution of the working time had a long-tail; 69.0\% of all the working times were below the average.
The hourly wage averaged \$9.15 (SD = 29.11; Median = 4.23).
Among all HIT records, 2,270 (29.8\%, in 674 HIT groups (42.4\%)) were above the U.S. minimum wage (7.25 USD).

For the final labels for the working time, workers chose TIME\_ALL for 3,802 (50.0\%) HIT records, TIME\_BTN for 2,525 (33.2\%) records, TIME\_FOCUS for 748 (9.8\%) records, and TIME\_CUSTOM for the rest 533 (7.0\%) records.
This partly supported our hypotheses, in that TIME\_FOCUS and TIME\_CUSTOM occurred less than the others, but went against our expectation that TIME\_ALL would occur more than TIME\_CUSTOM by 16.8 points.

We then conducted a follow-up analysis on TIME\_BTN.
Our investigation into the workers' button usage showed that the button was clicked at least once in 7,037 (92.5\%) HITs, that the usage per worker averaged 95.8\%, and that merely 16 workers (19.3\%) were below the average. 
The button appeared to be correctly utilized in most cases. However, the analysis also revealed that participants seldom stopped the timer.
The timer was stopped once in 73 HITs, twice in 8 HITs, and three times or more in 8 HIT,s while the timer was never stopped in the remaining 6,948 HITs.
We then analyzed the differences between TIME\_BTN and TIME\_ALL for microtasks which TIME\_BTN was chosen for the final choices.
The results showed that the differences in 72.5\% of the records were less than 5 s, and those in 85.1\% were less than 10 s.
The results indicate that most of the TIME\_BTN records that were chosen for the final labels were nearly equal to TIME\_ALL, (\textit{i.e.,} the button was immediately clicked when tasks started and finished without taking a break).
On the other hand, it was not possible for us to rigidly evaluate whether the button was correctly used, because our collected dataset did not contain tracking information of workers during HITs. We leave this as a topic for future work.

\subsection{TurkScanner: Hourly Wage Prediction}

TurkScanner predicts the hourly wages of microtasks in two steps: 1) estimate the working times of microtasks based on {\category HIT}, {\category WKR}, and {\category REQ}, in a machine learning-based approach; 2) calculate the hourly wage from the rewards and the estimated working times.

We predicted the working times of microtasks through regression using gradient boosted decision tree (GBDT) \cite{friedman2001}.
The model was trained with 148-dimensional feature vectors of task-relevant information (see Section 3.2), by minimizing the mean absolute error between the predicted and actual working times.
Upon training and testing, our GBDT model outputs the working times of microtasks on a \textit{log scale (base=10)}.
As shown in Figure \ref{fig:working_time_dist}, the working times of microtasks in our dataset admit a long-tail distribution. Taking the logarithm prevents the model from being excessively optimized for short-length microtasks and being affected by outliers.

The model evaluation was conducted through four-fold cross validation.
When partitioning the dataset, we picked 25\% of the 83 workers, and used all their HIT submission records for the test set. This means that the same worker never belonged to both the training and test sets.
Therefore, the validation results indicate the extent to which the model is capable of predicting the working time without being trained using HIT submission records of the same worker.
To obtain the predicted working times for all the microtasks in the dataset, we tested the model for all the validation pairs, and analyzed them all together.
To prevent the results from being too dependent in each trial, we iterated training and testing 50 times, and then calculated the average working time for each HIT record to obtain the results for subsequent analysis.

\section{Results and Discussion}

In this section, we describe the evaluation results of TurkScanner.
We first analyze feature importance, followed by analyses of the prediction accuracy for the working time and hourly wage.

\begin{figure}
\includegraphics[width=8.5cm]{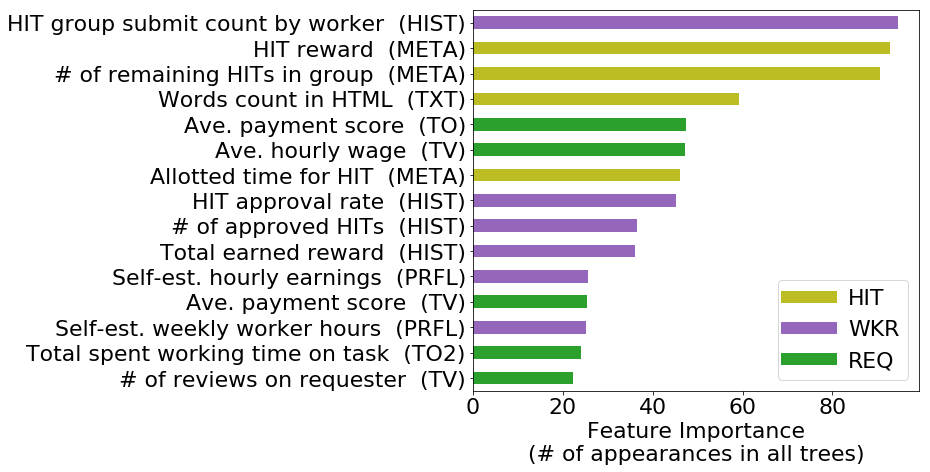}
\caption{Top-15 feature importance rankings.
Features for batch size, submission counts, and some explicit features (\textit{i.e.,} price, time, microtask contents) seem to implicate working time length.
Worker profiles are possibly effective, representing workers' proficiency on tasks.
}
\label{fig:features}
\end{figure}

\subsection{Feature Importance}

We first measured the feature importance, to better understand how each feature dimension contributes to predicting the working time.
Among the diverse methods available to measure the feature importance, we selected ``weight'' provided by XGBoost \cite{chen2016xgboost}, which counts how many times a feature is utilized for splitting across all generated trees.
We iterated the training of the initial model 50 times, and then took the averages of the feature importance values for the following analysis.

Figure \ref{fig:features} visualizes the importance ranking of the features for the working time prediction.
Features that could provide indications to estimate the microtask size appeared to be especially important.
For instance, larger HIT group submission count numbers (first place in the ranking), the number of remaining HITs in the group (third), and the word count in the HTML would imply how quickly HITs can be completed, because shorter HITs are more likely to be performed repeatedly.
More intuitively, features concerning the price (HIT reward: second), time (time allotted for a HIT: seventh), and task contents (word count in HTML: fourth) could also help to represent the working time.
These features also appeared to be effective as reputations by workers ({\category REQ} features: fifth, sixth, 12th, 14th, and 15th).
Worker profiles and task submission histories were also thought to be effective ({\category WKR} features: 8th--11th and 13th).
All these features represent workers' proficiencies with their working hours, accepted HITs, and earned rewards. This could possibly help to ensure the reliability of the working time prediction.


\begin{figure*}[t]
\includegraphics[width=15.5cm]{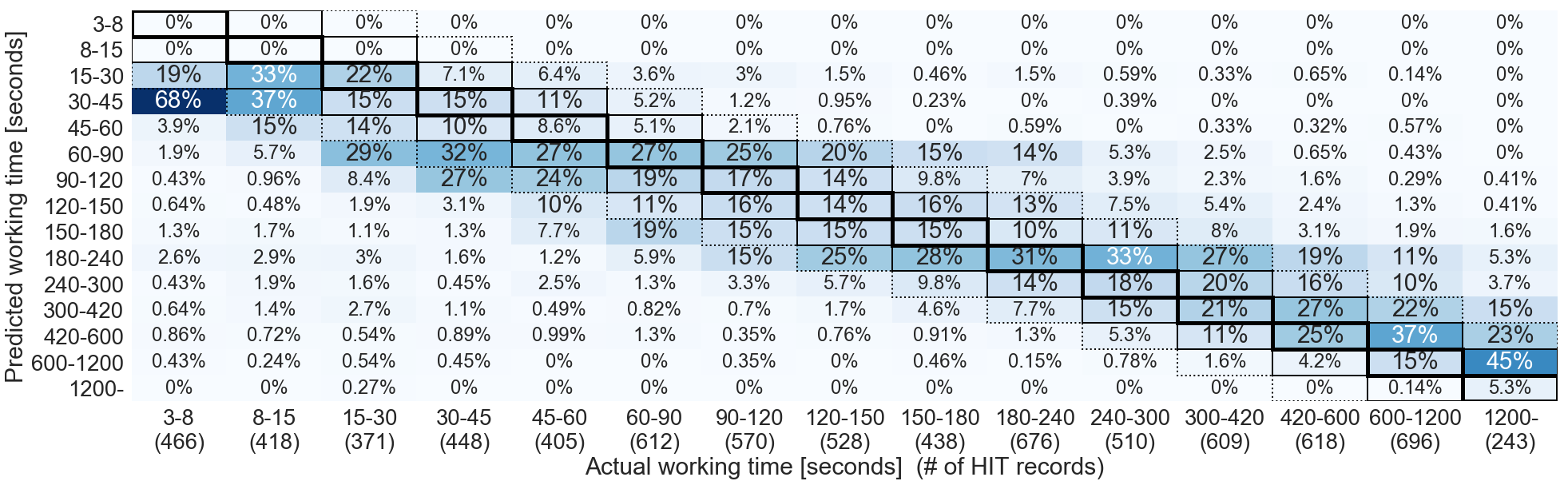}
\caption{
    Working time prediction results in a confusion matrix, illustrated by a heat map.
    A large portion of the prediction results are distributed diagonally, which implies that the model successfully captured the trend in the working time prediction.
}
\label{fig:heatmap_working_time}
\end{figure*}

\subsection{Working Time Prediction}


Figure \ref{fig:heatmap_working_time} presents a heat map of the confusion matrix representing the distribution of the working time prediction results per working time bin.
The size of each bin increases gradually towards the right, where the bin of the leftmost columns gathers all the HIT records whose actual working times are between 3 and 8 s, whereas that of the second right-most column includes those between 600 and 1,200 s.
Note that we only utilized this binning rule for the analysis: TurkScanner outputs consist of float values for the working time.

The heat map indicates that a large proportion of the predicted working times hit the bin or a nearby bin.
Seventeen percent of all the HIT submission records are in the diagonal cells in the heat map, surrounded by bold grid lines, ({\em i.e.}, they are categorized in the correct bins).
Allowing the prediction results to be categorized into neighboring bins, the proportion of correct predictions increases to 47.4\% with a one-cell difference (indicated by thin grid lines), and 70.8\% with a two-cell difference (indicated by dotted lines).
We also note that the predicted working times of HIT records with shorter working time labels (less than 60 s) are likely to be longer. Likewise, the predicted working times of HITs with longer working time labels (more than ~600 seconds) tend to be shorter.
This may be because the distribution of the logarithmic working time labels is closer to the normal distribution (see Figure ~\ref{fig:working_time_dist}b).
As mentioned in Section 3.4, the objective function of the GBDT algorithm trains the model such that the overall error across all the data is minimized.
Therefore, the model may have been trained to reduce the prediction error for HIT records with medium-length working times, where the largest amount of data samples are.



\subsection{Hourly Wage Calculation}

We calculated the hourly wage using the rewards and predicted working times of microtasks.
Over all the tested HIT records, the predicted hourly wage averaged 5.21 USD (SD = 4.53; Median = 4.20).
For N = 5,297 (69.6\% of all the collected HIT records) the hourly wage was predicted within a 75\% error, and for N = 6,412 (84.3\% of all the records) it was predicted within a 100\% error.

The prediction performed reasonably well for HIT records with actual hourly wages lower than around 15 USD. On the other hand, many of the HIT records with higher hourly wages were not predicted to be as high as they actually were.
To further analyze the incorrect results, we directly inspected the corresponding HIT records.
As a result, we determined that most of these were {\em (i)} survey HITs with external URL(s) and {\em (ii)} microtasks for which contents were dynamically rendered with JavaScript.
These two types of HITs are very similar in that they do not have much static HTML content by themselves.
Because we revealed in our feature analysis that some types of HTML content ({\em e.g.}, text counts in the HIT page, URL counts, and input tags) affected the prediction results, the model might have not been able to predict these HITs accurately.

\section{Limitations and Future Work}


\subsection{Limitations}

We first describe limitations concerning the data cleaning described in Section 3.3.
Our data cleaning method could be improved by scraping additional information, so that more noisy data are removed from the dataset.
First, tracking the statuses of completed microtasks (\textit{i.e.,} accepted or rejected) could filter out unreliable data.
If microtasks are rejected, then some of the scraped data are likely to be inconsistent. For instance, the working time may be too short compared to the task.
Second, we could track worker behavior such as cursor movements, scrolling, and the keyboard input for a certain time window.
Using such information, we can verify whether the history of recorded working times corresponds to the user behavior, and remove records if not.

We could also collect a more refined dataset by tracking microtask bonuses.
TurkScanner only considered fixed rewards for microtasks, and bonuses were not taken into account for the hourly wage prediction.
This would underestimate the value of microtasks whose reward schemes largely depend on bonuses for the microtask content.
Tracking the sizes of bonuses paid for each microtask would help TurkScanner to build a more accurate prediction model.

\subsection{Future Work}

Our next step will be to develop TurkScanner as a support tool for workers and requesters to solve real-world problems.
We will observe how TurkScanner could help workers reach better decisions on which microtasks to start.
We will also incorporate aspects that can be found in other similar recommender systems, such as predicting HIT acceptance rates, or recommending HITs based on workers' task preferences.
TurkScanner could also assist requesters with performing better price setting.
It could predict the hourly wage for microtasks prior to posting on the platform, and suggest that requesters increase their rewards for better wages.
This would greatly contribute to improving the average microtask price in the market, and accelerate future crowdsourcing research.

More generally, TurkScanner represents a new approach to understanding and predicting the times required for arbitrary user interface tasks.
This could be utilized in a variety of settings, such as setting wages in other domains, helping people to better organize or schedule their time, or in automated usability testing.
One can imagine a scenario in which every new task a government or company posts comes accompanied with an estimate of the cost of introducing the new task.
It may still be required to fill out an extra web form to justify conference travel, but at least it would be explicitly known how much time (and thus money) would be required to do so.

\section{Conclusion}

In this study, we tackled the challenge of predicting the hourly wages of microtasks based on data collected from previous workers.
We first presented a data collection method with our web browser extension for gathering data about crowd work and labeling the data with accurate working times.
We asked workers to select their answers from choices of working times recorded either automatically and manually by the workers themselves.
We then proposed TurkScanner, a system based on the GBDT regression model, to predict working times, and thus calculate hourly wages as its final output.
Our evaluation results indicated TurkScanner would need further improvement on its prediction performance. Nonetheless, we clearly showed the possibility that workers can know whether their crowd work will be worth the pay before actually embarking on it, which would make crowd work more transparent and beneficial for workers and requesters.

\bibliographystyle{ACM-Reference-Format}
\bibliography{main}

\end{document}